\pdfoutput=1
\documentclass[12pt]{article}

\usepackage{graphicx}

\input{epsf}

\def\rf#1{(\ref{eq:#1})}
\def\lab#1{\label{eq:#1}}

\def\br{\begin{eqnarray}}
\def\er{\end{eqnarray}}
\def\be{\begin{equation}}
\def\ee{\end{equation}}
\def\({\left(}
\def\){\right)}

\def\rlx{\relax\leavevmode}

\def\ve{\varepsilon}

\newcommand{\sbr}[2]{\left\lbrack\,{#1}\, ,\,{#2}\,\right\rbrack}

\def\IZ{\rlx\hbox{\sf Z\kern-.4em Z}}
\def\IR{\rlx\hbox{\rm I\kern-.18em R}}
\def\IC{\rlx\hbox{\,$\inbar\kern-.3em{\rm C}$}}
\def\one{\hbox{{1}\kern-.25em\hbox{l}}}

\begin{document}

\begin{titlepage}
\vspace*{-1cm}

\vskip 2cm

\vspace{.2in}
\begin{center}
{\large\bf A remark on the asymptotic form of BPS multi-dyon solutions and their conserved charges}
\end{center}

\vspace{.5cm}

\begin{center}
C. P. Constantinidis~$^{\dagger}$, L. A. Ferreira~$^{\star}$ and G. Luchini~$^{\dagger}$

\vspace{.3 in}
\small

\par \vskip .2in \noindent 
$^{\dagger}$ Departamento de F\'isica\\
 Universidade Federal do Esp\'irito Santo (UFES),\\
 CEP 29075-900, Vit\'oria-ES, Brazil
 
\par \vskip .2in \noindent 
$^{\star}$ Instituto de F\'\i sica de S\~ao Carlos; IFSC/USP;\\
Universidade de S\~ao Paulo  \\ 
Caixa Postal 369, CEP 13560-970, S\~ao Carlos-SP, Brazil\\

\vspace{2cm}

\normalsize
\end{center}

%\vspace{.5in}

\begin{abstract}

\noindent We evaluate the gauge invariant, dynamically conserved charges, recently obtained from the integral form of the Yang-Mills equations,  for the BPS multi-dyon solutions of a Yang-Mills-Higgs theory associated to any  compact semi-simple gauge group $G$. Those charges are shown to correspond to the eigenvalues of the next-to-leading term of the asymptotic form of the Higgs field at spatial infinity, and so coinciding with the usual topological charges of those solutions.  Such results show that  many of the topological charges considered in the literature are in fact dynamical  charges, which conservation follows from the global properties of classical Yang-Mills theories encoded into their  integral  dynamical equations. The conservation of those charges can not be obtained from the differential form of Yang-Mills equations.

\end{abstract} 
\end{titlepage}

\section{Introduction and Conclusions}
\label{sec:intro}
\setcounter{equation}{0}

The purpose of this paper is to evaluate the gauge invariant, dynamically conserved charges, proposed in \cite{ym1,ym2}, for the BPS multi-dyon solutions of a Yang-Mills-Higgs theory associated to any  compact semi-simple gauge group $G$. Such conserved charges are obtained from the integral equations of Yang-Mills theory and can be obtained either from a volume ordered integral on the whole tridimensional space, or as a surface ordered integral on its border, which is the two-sphere $S^2_{\infty}$ at spatial infinity. The latter form is simpler to evaluate in general, and the conserved charges correspond to the eigenvalues of the operator
\be
Q=P_2 e^{ie\int_{S^2_{\infty}}d\tau d\sigma\, W^{-1}\, (\alpha F_{\mu\nu}+\beta {\widetilde F}_{\mu\nu}) \,W\,\frac{dx^{\mu}}{d\sigma}\frac{dx^{\nu}}{d\tau}}
\lab{charge}
\ee
where $P_2$ means surface ordered integration as we explain below, $F_{\mu\nu}=\partial_{\mu}A_{\nu}-\partial_{\nu}A_{\mu}+i\,e\,\sbr{A_{\mu}}{A_{\nu}}$, is the field tensor and ${\widetilde F}_{\mu\nu}\equiv \frac{1}{2}\,\varepsilon_{\mu\nu\rho\lambda}\, F^{\rho\lambda}$, its Hodge dual. $W$ is the Wilson line, i.e. the holonomy  of the connection $A_{\mu}$ along paths scanning $S^2_{\infty}$, and $\alpha$ and $\beta$ are arbitrary parameters. Such charges are conserved in time for any solution of the Yang-Mills equations in the presence of sources, as long as the following boundary conditions are satisfied: $F_{\mu\nu} \rightarrow 1/r^{3/2 + \delta}$ and $J_{\mu}\rightarrow 1/r^{2+\delta^{\prime}}$, with $\delta\, , \, \delta^{\prime} >0$, for $r\rightarrow \infty$, $r$ being the radial distance, and $J_{\mu}$ is the current associated with the external fields like fermions, Higgs fields, etc (see \cite{ym1,ym2} for details). 
In this paper we are concerned with time-independent solutions of the BPS equations \cite{bogo,prasad,goddardolivereview,mantonsut,weinbergyi,weinbergbook,sutcliffereview,shnir}
\be
B_i= \cos \gamma\; D_i \phi \qquad \qquad \qquad E_i= \sin \gamma\; D_i \phi \qquad \qquad \qquad D_0 \phi =0
\lab{bpseq}
\ee
where $B_i \equiv -\frac{1}{2}\,\varepsilon_{ijk}\,F_{jk}$ and $E_i\equiv F_{0i}$, $i,j,k=1,2,3$, are the non-abelian magnetic and electric fields respectively,  $\phi$ is the Higgs field transforming under the adjoint representation of the gauge group $G$, $D_i=\partial_i*+i\,e\,\sbr{A_{i}}{*}$, and $\gamma$ is an arbitrary angle. Since we are considering time-independent solutions, the last two equations of \rf{bpseq} can be solved by taking
\be
A_0= -\sin \gamma\; \phi
\ee
The first equation in \rf{bpseq} has been studied extensively and it can be solved exactly through a variety of techniques. Two and multi-monopole solutions have been constructed using  the Nahm transform (ADHMN)  \cite{nahm1,nahm2,nahm3,nahm4}, twistors  
\cite{ward1,ward2,corrigangoddard,prasadsinha,prasadrossi}, Backl\"und transformations \cite{forgacs1,forgacs2,forgacs3},  rational maps \cite{hitchin1,donaldson,hurtubise1}, and other direct methods \cite{jutta,betti} .  In addition, many exact properties  of BPS monopoles have been obtained 
\cite{jaffe,hitchin2,hurtubise2,hurtubisemurray,nogradi,dereknogradi,mantonsut,weinbergyi,weinbergbook}. 

In order to evaluate the charges \rf{charge} one needs the  analytic expressions for the asymptotic form of the gauge fields at spatial infinity. Despite the many developments on BPS monopoles and dyons, the asymptotic form of the gauge fields is not easy to extract from the known exact solutions. The asymptotic form of the Higgs field is better studied, specially its modulus \cite{hurtubise2,hurtubisemurray,dereknogradi}, but that  does not help much on the evaluation of the charges \rf{charge}. A result which is really relevant is the so-called {\em generalised inverse square law} where the asymptotic form of the magnetic field is assumed to be  \cite{goddardnuytsolive,goddardolivereview}
\be
B_i = \frac{1}{e}\,\frac{{\hat r}_i}{r^2}\, g\({\hat r}\) \qquad\qquad\qquad\qquad D_i\, g\({\hat r}\)=0 \qquad\qquad\qquad \qquad r\rightarrow \infty
\lab{geninvsqlaw}
\ee
with ${\hat r}={\vec r}/r$ being the unit vector in the radial direction, and $g\({\hat r}\)$ being an element of the Lie algebra of the gauge group $G$. Obviously the first equation in \rf{geninvsqlaw} is what characterizes a monopole or multi-monopole solution, but the second equation is an assumption which even though seems to be true for large classes of known monopole solutions, has not been proven in general. Indeed, in the literature the equations \rf{geninvsqlaw} are used  to calculate the magnetic charges  as topological charges through homotopy arguments, and they are shown to be related to the co-weights of the little group $H$ to which the gauge  group $G$ has been spontaneously broken \cite{goddardnuytsolive,goddardolivereview,hurtubisemurray,weinberg1,weinberg2,weinbergyi,weinbergbook}.  In addition, it is known  that for large classes of  BPS multi-monopole solutions  the Higgs field at spatial infinity behaves as 
\be
\phi = \phi_0\({\hat r}\) + \frac{\phi_1\({\hat r}\)}{r} + \ldots 
\qquad\qquad\qquad \qquad r\rightarrow \infty
\lab{higgsatinfinity}
\ee
where $\phi_0\({\hat r}\)$ and $\phi_1\({\hat r}\)$ are Lie algebra elements of the gauge group $G$, and where the   remaining terms decay   algebraically and exponentially with $r$. 

Instead of trying to extract the asymptotic behavior of the gauge fields (at spatial infinity) from the  known exact solutions, we shall assume an asymptotic ansatz for them, which is compatible with all known facts about dyons solutions, and then impose the BPS equation \rf{bpseq} asymptotically. We shall work in a gauge where
\be
{\hat r}\cdot {\vec A}=0
\lab{radialgauge}
\ee
and assume the following asymptotic form for the gauge fields
\be
A_i=-\frac{1}{e}\, \varepsilon_{ijk}\, {\hat r}_j\,\left[\frac{K_k\({\hat r}\)}{r}+\frac{L_k\({\hat r}\)}{r^2}+\ldots \right]
\qquad \qquad \qquad r\rightarrow \infty
\lab{asymptoticgauge}
\ee
with $i,j,k=1,2,3$, and $K_i\({\hat r}\)$ and $L_i\({\hat r}\)$ being Lie algebra elements depending only on the radial direction ${\hat r}$, but not on the radial distance $r$. The remaining terms in \rf{asymptoticgauge} decay algebraically and exponentially  as $r\rightarrow \infty$. The results we obtain can be summarized as follows:
\begin{enumerate}
\item The covariant derivative of the leading term of the Higgs field in \rf{higgsatinfinity} decays faster than $1/r$, i.e. 
\be
r\, D_i\, \phi_0 \rightarrow 0 \qquad\qquad\qquad {\rm as} \qquad\qquad r\rightarrow \infty
\lab{covderphi0result}
\ee
\item The leading and next to leading terms in  \rf{higgsatinfinity} commute, i.e.
\be
\sbr{\phi_0}{\phi_1}=0
\lab{niceconsequence}
\ee
and so, $\phi_1$ belongs to the Lie algebra of the little group $H$ to which the gauge  group $G$ has been spontaneously broken.
\item For a compact and semisimple gauge group $G$  it is quite reasonable to assume that $\phi_0$ is a semisimple element of the Lie algebra ${\cal G}$ of $G$, in the sense that its adjoint action splits ${\cal G}$ into image and kernel without any intersection. More precisely, we assume that 
\be
{\cal G} = {\rm Ker}_{{\rm adj}\phi_0} + {\rm Im}_{{\rm adj}\phi_0}
\lab{phi0semisimple1}
\ee
with
\be
\sbr{\phi_0}{{\rm Ker}_{{\rm adj}\phi_0}}=0 \qquad\qquad \qquad 
{\rm Im}_{{\rm adj}\phi_0}=\sbr{\phi_0}{{\cal G}}\qquad\qquad \qquad 
{\rm Ker}_{{\rm adj}\phi_0} \cap {\rm Im}_{{\rm adj}\phi_0} = 0
\lab{phi0semisimple2}
\ee
If that is so, then it follows that the covariant derivative of the next to leading term of the Higgs field in \rf{higgsatinfinity} also decays faster than $1/r$, i.e. 
\be
r\, D_i\, \phi_1 \rightarrow 0 \qquad\qquad\qquad {\rm as} \qquad\qquad r\rightarrow \infty
\lab{phi1covconst}
\ee
As a consequence of \rf{covderphi0result} and \rf{phi1covconst}, it follows that 
\be
D_i\phi= - \frac{ {\hat r}_i}{r^2}\, \, \phi_1\({\hat r}\)
+ O\(\frac{1}{r^3}\)
\lab{covderhiggsexpfinal}
\ee
\item The BPS equation \rf{bpseq} implies that $g\({\hat r}\)$ defined in \rf{geninvsqlaw} must be given by 
\be
g\({\hat r}\)=-e\,\cos\gamma\,\,\phi_1\({\hat r}\)
\lab{gphi1rel}
\ee
Therefore, the second condition in \rf{geninvsqlaw} follows from  \rf{phi1covconst}. Note that the BPS equations  \rf{bpseq} alone are not sufficient to imply the generalized inverse square law  \rf{geninvsqlaw}. One needs to assume that $\phi_0$ is a semisimple element of ${\cal G}$.
\item The fact that $\phi_0$ is a semisimple element of ${\cal G}$, as defined in \rf{phi0semisimple1} and \rf{phi0semisimple2}, implies that the surface ordering $P_2$ is not necessary in the evaluation of the dynamically conserved charges \rf{charge}, and the operator $Q$ given in  \rf{charge} becomes
\be
Q = e^{-i\,4\,\pi\,e\,\(\alpha \, Q_B+ \beta\, Q_E\)}
\lab{qexponential}
\ee
where
\br
Q_B&=& \frac{1}{4\,\pi}\,\int_{S^2_{\infty}}\, d\Sigma_i \, W^{-1}\, B_i\,W
 = -\cos \gamma\; \phi_1\({\hat r}_R\)=\frac{g\({\hat r}_R\)}{e}
\lab{finalmagneticcharge} \\
Q_E&=& \frac{1}{4\,\pi}\,\int_{S^2_{\infty}}\, d\Sigma_i \, W^{-1}\, E_i\,W 
= -\sin \gamma\; \phi_1\({\hat r}_R\)=\tan \gamma\, \frac{g\({\hat r}_R\)}{e}
\lab{finalelectriccharge} 
\er
where the integration is over the two-sphere $S^2_{\infty}$ at spatial infinity, and $\phi_1\({\hat r}_R\)$ is the value of the Lie algebra element $\phi_1$, defined in \rf{higgsatinfinity}, on an arbitrarily chosen point ${\hat r}_R$ of $S^2_{\infty}$. The physical magnetic  and  electric charges, which are dynamically conserved, are the eigenvalues of the operators $Q_B$ and $Q_E$ respectively, and those are determined by the eigenvalues of the operator $\phi_1\({\hat r}_R\)$. Such eigenvalues do not depend upon the choice of the point ${\hat r}_R$ of $S^2_{\infty}$ (see \cite{ym1,ym2}).

\end{enumerate}

As it is well known the  charges of magnetic monopole solutions in a gauge theory with symmetry spontaneously broken by the Higgs mechanism are topological charges determined by the second homotopy group of the Higgs vacua. The magnetic and electric charges we have evaluated here are dynamically conserved, and so are not of a topological nature. They are conserved in fact for any solution of the Yang-Mills equations that satisfy the boundary conditions discussed above.  However, the values of the magnetic charges \rf{finalmagneticcharge} coincide with those obtained by topological methods. Indeed, the eigenvalues of  $\phi_1\({\hat r}_R\)$, and so of  $g\({\hat r}_R\)$ (see \rf{gphi1rel}), must relate to the integers appearing in the so-called {\em skyline diagram} used in the ADHMN construction \cite{hurtubisemurray,houghton,weinbergyi}. In addition, it is known that the modulus of $\phi_1$ gives the monopole number \cite{hitchin2,hurtubise2}. Perhaps the best way of understanding the relation among our dynamical magnetic charges and the topological ones is through the homotopy considerations of  \cite{goddardnuytsolive,goddardolivereview}. Indeed, those authors have shown that the homotopy classes, and so the magnetic charges, are related to path dependent phase factors which are in fact  elements  of the little group $H$ to which the gauge group $G$ has been spontaneously broken. Such elements can be evaluated using the non-abelian Stokes theorem and the magnetic charges can be expressed in the same form as in  \rf{finalmagneticcharge}. They do not obtain however the electric charges.  Indeed, to construct the charges \rf{charge} and show their conservation one has to use a higher non-abelian Stokes theorem involving a two-form connection \cite{ym1,ym2}, and not just a one-form connection like in the usual Stokes theorem. 

Note that using either the usual non-abelian Stokes theorem or its higher version involving a two-form connection (see next section for details) one observes that the operator $Q$ has to be unity for $\alpha=1$ and $\beta=0$, i.e. (see \rf{qexponential}, \rf{finalmagneticcharge}  and \rf{gphi1rel})
\be
Q\(\alpha=1\,,\,\beta=0\)= e^{-i\,4\,\pi\,e\, Q_B}=e^{-i\,4\,\pi\,g\({\hat r}_R\)} =\one
\lab{magneticquantum}
\ee
and that is the usual quantization of the magnetic charges for monopole solutions \cite{goddardnuytsolive,goddardolivereview}. 

Another point to stress is that the charge operator $Q$ given in \rf{charge} depends  on two arbitrary parameters $\alpha$ and $\beta$. If one expands $Q$ as a power series in those parameters (see \rf{chargeexpand}), it follows from our construction that each coefficient in such expansion is an independent conserved charge, and so one obtains an infinite number of non-local conserved charges involving higher powers of the  field tensor and its Hodge dual. It is not clear yet the physical relevance of such infinity of charges. In the case of BPS multi-dyon solutions discussed in this paper, the condition \rf{phi1covconst} implies that the higher charges are in fact powers of the first order magnetic and electric charges, and so $Q$ becomes an ordinary exponential as given in \rf{qexponential}. The same charges were evaluated in \cite{ym1,ym2} for the $SU(2)$ 'tHooft-Polyakov monopole and Julia-Zee dyon, as well as for the Wu-Yang monopole, and the same thing happens, i.e. the higher charges are powers of the first ones. Again, the mathematical reason for that is the fact that the Lie algebra elements determining the magnetic and electric charges are covariantly constant at spatial infinity. It would be interesting to investigate if there is a no-go theorem involving the infinity of charges coming from the expansion of the operator $Q$, given in \rf{charge}, in powers of $\alpha$ and $\beta$. Perhaps, extending the calculations of the present paper to non-BPS monopoles and dyons for higher gauge groups, to show that they also have only a finite number of charges, would clarify such point. 

We point out that the charges \rf{charge} differ from the usual dynamically conserved electric and magnetic charges discussed in the literature. Such charges can be obtained directly from the differential form of Yang-Mills equations or through the Noether theorem, and are given by  
\br
Q_B^{YM}= \frac{1}{4\,\pi}\,\int_{S^2_{\infty}}\, d\Sigma_i \,  B_i 
\qquad\qquad\qquad\qquad\qquad 
Q_E^{YM}= \frac{1}{4\,\pi}\,\int_{S^2_{\infty}}\, d\Sigma_i \,  E_i 
\lab{ymcharge} 
\er
As pointed out already by C.N. Yang and R. Mills in their original paper \cite{ymoriginal}, as well as in many text-books (see for instance \cite{sweinbergbook}), the charges \rf{ymcharge} are not really gauge invariant. They are invariant only under gauge transformations that go to a constant at spatial infinity. In addition, they vanish when  evaluated  in the 'tHooft-Polyakov or Wu-Yang monopole solutions, in the Julia-Zee dyon \cite{wuyang,shnir}, and perhaps in many other monopole solutions.

We also point out that our charges \rf{finalmagneticcharge} relate to the topological charges of magnetic monopoles used in the literature. Indeed, for BPS monopoles such topological charges are  given by \cite{bogo,prasad,goddardolivereview,mantonsut,weinbergyi,weinbergbook,sutcliffereview,shnir}
\be
Q_B^{{\rm Top.}}= \frac{1}{2}\,\int_{\IR^3} d^3r \, {\rm Tr}\(B_i\,D_i\phi\)
= \frac{1}{2}\,\int_{S^2_{\infty}} d\Sigma_i \, {\rm Tr}\(B_i\,\phi\)
= \frac{1}{2\, e}\,\int_{S^2_{\infty}} d\Omega \, {\rm Tr}\(g\({\hat r}\)\,\phi_0\({\hat r}\)\)
\lab{topcharge}
\ee
where we have used \rf{geninvsqlaw} and \rf{higgsatinfinity}, and where $d\Sigma_i={\hat r}_i\,r^2\,d\Omega$, with $d\Omega$ being the solid angle element. Through a gauge transformation, we can rotate the $\phi_0$-component of Higgs field defined in \rf{higgsatinfinity}, at the reference point ${\hat r}_R$, into the Cartan sub-algebra of the gauge group $G$, i.e.
\be
\phi_0\({\hat r}_R\)= {\vec v}\cdot {\vec h}
\ee
where $h_j$ are the elements of the Cartan-Weyl basis of the  Cartan sub-algebra of $G$, normalized as ${\rm Tr}\(h_j\,h_k\) = \delta_{j\,k}$. Therefore, the modulus of the vector ${\vec v}$ is the v.e.v. of the Higgs field, i.e. $\phi_0^2={\vec v}^2$. We can now perform a gauge transformation of the unbroken gauge group $H\subset G$, which leaves $\phi_0$ invariant, to rotate the next to leading term $\phi_1$ of the Higgs field (see \rf{higgsatinfinity}), at the reference point ${\hat r}_R$, into the Cartan sub-algebra of the gauge group $G$ as well. From \rf{gphi1rel} one then gets that
\be
g\({\hat r}_R\)={\vec \omega}\cdot {\vec h}
\lab{gcartan}
\ee
The quantization condition \rf{magneticquantum} however implies that 
\cite{goddardnuytsolive,englert,corriganolive}
\be
{\vec \omega} = \sum_{a=1}^{{\rm rank}\,G} n_a\,\frac{{\vec \alpha}_a}{\alpha_a^2}
\lab{omegaquant}
\ee
where $n_a$ are integers, and ${\vec \alpha}_a$ are the simple roots of the Lie algebra of the gauge group $G$. The reason for that is that the elements of Chevalley basis of the Cartan sub-algebra of $G$, namely $2 \,{\vec \alpha}_a\cdot {\vec h}/\alpha_a^2$, have integer eigenvalues in any finite dimensional representation of the gauge group $G$. It can then be shown that the topological charge \rf{topcharge} becomes
\be
Q_B^{{\rm Top.}}= \frac{2\,\pi}{e}\, {\vec v}\cdot {\vec \omega}
=\frac{2\,\pi}{e}\,  \sum_{a=1}^{{\rm rank}\,G} n_a\,\frac{{\vec \alpha}_a\cdot {\vec v}}{\alpha_a^2} 
\lab{topcahragecartan}
\ee
If the vector ${\vec v}$ is not orthogonal to any simple root ${\vec \alpha}_a$, then the gauge group $G$ is broken to $H=\left[U(1)\right]^{{\rm rank}G}$. Therefore, in this case the topological charge depends on all {\em magnetic weights} $n_a$. However, if ${\vec v}$ is orthogonal to $m$ of the simple roots then $G$ is broken to $H=K\otimes \left[U(1)\right]^{{\rm rank}G-m}$, where $K$ is a non-abelian group which simple roots are the $m$ simple roots of $G$ otrthogonal to ${\vec v}$. In such a case $m$ of the {\em magnetic weights} $n_a$ do not contribute to the topological charge. 

Notice that, since ${\vec v}$ is a vector in the root space of the Lie algebra of $G$, it can be expanded in terms of the fundamental weights ${\vec \lambda}_b$ (satisfying $2\,{\vec \lambda}_b\cdot {\vec \alpha}_a/\alpha_a^2=\delta_{a\,b}$), as ${\vec v}=\sum_{a=1}^{{\rm rank}\,G} v_a\, {\vec \lambda}_a$.  Therefore 
$Q_B^{{\rm Top.}}=\frac{\pi}{e}\,  \sum_{a=1}^{{\rm rank}\,G} n_a\,v_a$.

As we have shown above, the dynamically conserved charges we have constructed, are the eigenvalues of the operator $Q_B$, and so of $Q_E$, given in \rf{finalmagneticcharge}  and \rf{finalelectriccharge}. Therefore, our charges are given by 
\be
q_B=\mbox{\rm eigenvalues of $Q_B$}= \frac{1}{2\,e}\,\sum_{a=1}^{{\rm rank}\,G} n_a\,
\left( m_a^{(1)}\, ,\, m_a^{(2)}\, ,\,\ldots \, ,\,m_a^{(d)}\right)
\lab{dinamicalchargesexplicit}
\ee
where
\be
m_a^{(s)}=\mbox{\rm eigenvalues of $2{\vec \alpha}_a\cdot {\vec h}/\alpha_a^2$}
\ee
and $d$ is the dimension of the representation of the gauge group where the eigenvalues are being evaluated. Note that $m_a^{(s)}$ are integers in any finite dimensional representation of the gauge group, since they are the eigenvalues of the elements of the Chevalley basis of the Cartan sub-algebra of $G$. Therefore such charges satisfy the quantization condition \rf{magneticquantum} in any finite representation. So, our charges are vectors in a given representation, and do not involve a projection onto the v.e.v. ${\vec v}$ of the Higgs field. So, unlike the topological charge \rf{topcharge}, our dynamically conserved charges do not miss some of the {\em magnetic weights} $n_a$ , when there is a non-abelian factor $K$ in the unbroken gauge group $H$. The missing of such magnetic weights has been an issue in the literature (see for instance \cite{weinbergbook,shnir}). Notice that the integers $m_a^{(s)}$ are fixed by the choice of the representation. If we change the representation the size of the vector $q_B$ changes, as well as its entries. However, they will be different combinations of the same magnetic weights $n_a$. Consequently, the dynamically conserved charges are essentially the magnetic weights. Even though they are widely used in the literature, they were never shown to be conserved either dynamically or topologically.

The charges \rf{charge} not only solve the problem of the non gauge invariance of \rf{ymcharge}, but also shows that the topological charges discussed in the literature can be expressed in terms of dynamically conserved charges. So, they unify in a simple and elegant way many facets and properties of classical Yang-Mills solutions. Even though the magnetic charges of monopoles and dyons have been constructed and studied as topological charges, the electric and magnetic charges of any Yang-Mills solution are dynamically conserved and gauge invariant, i.e. the eigenvalues of the operator \rf{charge}. Such property of classical Yang-Mills theories can not be disclosed  by the differential form of the Yang-Mills equations. They need the integral form of those equations. 

\section{The detailed arguments and calculations}
\label{sec:calc}
\setcounter{equation}{0}

As we mentioned in the introduction the charge operator $Q$ given in \rf{charge} can be calculated either by a surface ordered integral or a volume ordered integral. The two forms of it are given by 
\be
Q=P_2\, e^{ie\int_{S^2_{\infty}}d\tau d\sigma\, W^{-1}\, (\alpha F_{\mu\nu}+\beta {\widetilde F}_{\mu\nu}) \,W\,\frac{dx^{\mu}}{d\sigma}\frac{dx^{\nu}}{d\tau}}=  P_3\, e^{\int_{R^3} d\zeta d\tau  V{\cal J}V^{-1}}
\lab{chargefull}
\ee
where $R^3$ is the spatial sub-manifold of four dimensional Minkowski  space-time, and $S^2_{\infty}$ is the two-sphere at spatial infinity. The very same formulas  apply to curved space-time but we shall restrict ourselves to flat Minkowski  space-time (see \cite{ym1,ym2} for details). $P_2$ and $P_3$ mean surface and volume ordered integrations respectively, as we now explain. We choose a reference point 
$x_R$ on  $S^2_{\infty}$ and scan $R^3$ with closed two-dimensional surfaces, labelled by $\zeta$, based on $x_R$, such that $\zeta=0$ corresponds to the infinitesimal surface around $x_R$, and $\zeta=2\pi$ corresponds to $S^2_{\infty}$. We then scan each one of those two-dimensional surfaces with loops, labelled by $\tau$, starting and ending at $x_R$, such that $\tau=0$ corresponds to the infinitesimal loop around $x_R$, and $\tau=2\pi$ corresponds to another infinitesimal loop around the other side of $x_R$, obtained after the loops have spanned the whole surface. Each loop is paremeterized by $\sigma$, such that $\sigma=0$ and $\sigma=2\pi$ correspond to its initial and final points respectively, both at $x_R$ (see \cite{ym1,ym2} for details).

The  Wilson line $W$ is defined on any curve, parameterized by $\sigma$, through the equation
\be
\frac{d\,W}{d\,\sigma}+   i\,e\,A_{\mu}\,\frac{d\,x^{\mu}}{d\,\sigma}\,W=0
\lab{eqforw}
\ee
where $x^{\mu}$ ($\mu=0,1,2,3$) are the coordinates on the  four dimensional space-time. The quantity $V$ is defined on any surface through the equation
\be
\frac{d\,V}{d\,\tau}-V\, T\(A,\tau\)=0
\lab{eqforv}
\ee
with 
\be
T\(A,\tau\)\equiv ie\,
 \int_{0}^{2\pi}d\sigma W^{-1}\left[ \alpha F_{\mu\nu}+\beta{\widetilde F}_{\mu\nu}\right] W \frac{dx^{\mu}}{d\sigma}\frac{dx^{\nu}}{d\tau}
 \lab{tdef}
 \ee
In addition, ${\cal J}$, appearing in \rf{chargefull}, is given by  
\br
{\cal J}&=&
\int_0^{2\pi}d\sigma\left\{ ie\beta {\widetilde J}_{\mu\nu\lambda}^W
\frac{dx^{\mu}}{d\sigma}\frac{dx^{\nu}}{d\tau}
\frac{dx^{\lambda}}{d\zeta} \right. \nonumber\\
&+& \left. e^2\int_0^{\sigma}d\sigma^{\prime}
\sbr{\(\(\alpha-1\) F_{\kappa\rho}^W+\beta {\widetilde F}_{\kappa\rho}^W\)\(\sigma^{\prime}\)}
{\(\alpha F_{\mu\nu}^W+\beta {\widetilde F}_{\mu\nu}^W\)\(\sigma\)} \right.
\nonumber\\
&&\left. \times
\, \frac{d\,x^{\kappa}}{d\,\sigma^{\prime}}\frac{d\,x^{\mu}}{d\,\sigma}
\(\frac{d\,x^{\rho}\(\sigma^{\prime}\)}{d\,\tau}\frac{d\,x^{\nu}\(\sigma\)}{d\,\zeta}
-\frac{d\,x^{\rho}\(\sigma^{\prime}\)}{d\,\zeta}\frac{d\,x^{\nu}\(\sigma\)}{d\,\tau}\)\right\}
\lab{caljdef}
\er  
where ${\widetilde J}_{\mu\nu\lambda}$ is the Hodge dual of the matter current, i.e. $J^{\mu}=\frac{1}{3!}\varepsilon^{\mu\nu\rho\lambda}\,{\widetilde J}_{\nu\rho\lambda}$. The matter current may be due the Higgs fields, spinor matter fields, or any other fields carrying charges that couple to the Yang-Mills field $A_{\mu}$. To simplify the formulas we have used the notation $X^W\equiv W^{-1}\,X\,W$, with $W$ being the  Wilson line, and $X$ standing for the field tensor, its Hodge dual, or the dual of the matter currents. 

The integral Yang-Mills equations correspond to the equality between the volume and surface integrals in \rf{chargefull} on any three-dimensional volume $\Omega$ and its border $\partial \Omega$ \cite{ym1,ym2}. Then by considering a closed three-dimensional volume $\Omega_c$ (without border), it follows that the surface integral should be unity and so one gets $P_3\, e^{\int_{\Omega_c} d\zeta d\tau  V{\cal J}V^{-1}}=\one$. That statement is a conservation law in a way very similar to that used in integrable field theories. In fact it follows that ${\cal A}=\int d\tau  V{\cal J}V^{-1}$, is a flat connection in generalized loop space. That conservation law implies that, after  imposing the boundary conditions 
\be
J_{\mu} \sim \frac{1}{r^{2+\delta}}
\qquad \qquad{\rm and} \qquad  \qquad
F_{\mu\nu}\sim \frac{1}{r^{\frac{3}{2}+\delta^{\prime}}}\qquad \qquad {\rm for} \qquad\qquad  r\rightarrow \infty
\lab{boundcond}
\ee
with $\delta,\delta^{\prime}>0$,  the operator $Q$ has and iso-spectral evolution, i.e.
\be
Q\(t\)= U\(t\)\,Q\(0\)\,U^{-1}\(t\)
\lab{isospectral}
\ee
and so its eigenvalues are conserved in time (see \cite{ym1,ym2} for details). 

Note that the parameters $\alpha$ and $\beta$ are arbitrary in all those formulas. Therefore, we can expand $Q$ in a power series in those parameters to get (using the surface integral form of $Q$, and so integrating \rf{eqforv})
\br
Q&=& \one + \alpha\, \int_0^{2\pi} d\tau\, S\(\tau\)+ \beta\, \int_0^{2\pi} d\tau\, {\widetilde S}\(\tau\)
\nonumber\\
&+& \alpha^2\, \int_0^{2\pi} d\tau\, \int_0^{\tau} d\tau^{\prime}\, S\(\tau^{\prime}\)\, S\(\tau\)
+ \beta^2\, \int_0^{2\pi} d\tau\, \int_0^{\tau} d\tau^{\prime}\, {\widetilde S}\(\tau^{\prime}\)\, {\widetilde S}\(\tau\)
\nonumber\\
&+&\alpha\, \beta\, \int_0^{2\pi} d\tau\, \int_0^{\tau} d\tau^{\prime}\, \left[{\widetilde S}\(\tau^{\prime}\)\, S\(\tau\)+S\(\tau^{\prime}\)\, {\widetilde S}\(\tau\)\right] +\ldots
\lab{chargeexpand}
\er
with
\be
S\(\tau\)\equiv ie\,
 \int_{0}^{2\pi}d\sigma W^{-1}\, F_{\mu\nu}\, W \frac{dx^{\mu}}{d\sigma}\frac{dx^{\nu}}{d\tau}\; ; \qquad\quad
{\widetilde S}\(\tau\)\equiv ie\,
 \int_{0}^{2\pi}d\sigma W^{-1}\,{\widetilde F}_{\mu\nu}\, W \frac{dx^{\mu}}{d\sigma}\frac{dx^{\nu}}{d\tau}
 \lab{sstildedef}
 \ee
Therefore, from \rf{isospectral} and the independency of the parameters we should get that each term of the series is independently conserved, leading to an infinity of conserved charges.   

Note that if we take $\alpha=1$ and $\beta=0$ one gets that the quantity ${\cal J}$ in \rf{caljdef} vanishes. Therefore, \rf{chargefull} implies that
\be
P_2\, e^{ie\int_{S^2_{\infty}}d\tau d\sigma\, W^{-1}\, F_{\mu\nu} \,W\,\frac{dx^{\mu}}{d\sigma}\frac{dx^{\nu}}{d\tau}}=\one
\lab{bianchinaive}
\ee
and that is the equation that leads to the quantization of the magnetic charges as given in \rf{magneticquantum}. Such relation could also have been obtained by the usual non-abelian Stokes theorem for one-form connection, and that is how \rf{magneticquantum} was obtained in \cite{goddardnuytsolive,goddardolivereview}. One could interpret \rf{bianchinaive} as the integral form of the Bianchi identity. However, the true integral Bianchi identity is obtained by setting $\beta=0$ and leaving $\alpha$ arbitrary. One then gets from the integral Yang-Mills equations that
\be
P_2\, e^{ie\,\alpha\,\int_{\partial\Omega}d\tau d\sigma\, W^{-1}\,  F_{\mu\nu} \,W\,\frac{dx^{\mu}}{d\sigma}\frac{dx^{\nu}}{d\tau}}=  P_3\, e^{\alpha\(\alpha-1\)\,\int_{\Omega} d\zeta d\tau  V{\cal K}V^{-1}}
\lab{truebianchi}
\ee
with $\Omega$ being any three-dimensional volume and $\partial\Omega$ its border, and where 
\br
{\cal K}= e^2
\int_0^{2\pi}d\sigma 
 \int_0^{\sigma}d\sigma^{\prime}
\sbr{ F_{\kappa\rho}^W\(\sigma^{\prime}\)}{ F_{\mu\nu}^W\(\sigma\)} 
 \frac{dx^{\kappa}}{d\sigma^{\prime}}\frac{dx^{\mu}}{d\sigma}
\(\frac{dx^{\rho}\(\sigma^{\prime}\)}{d\tau}\frac{dx^{\nu}\(\sigma\)}{d\zeta}
-\frac{dx^{\rho}\(\sigma^{\prime}\)}{d\zeta}\frac{dx^{\nu}\(\sigma\)}{d\tau}\)
\nonumber
\er  
Of course, one can expand \rf{truebianchi} in powers of $\alpha$, and each coefficient on one side of the equation should equal the corresponding coefficient on the other side. We have checked that equation up to order $\alpha^2$ for the exact  $SU(2)$ BPS one-monopole solution and it is satisfied exactly \cite{new}. 

Let us now evaluate the charges for the BPS dyon solutions assuming the asymptotic form the of the Higgs and gauge fields as given in  \rf{higgsatinfinity} and \rf{asymptoticgauge} respectively. Using  \rf{asymptoticgauge} one gets that the magnetic field is given by
\be
{\vec B}= \frac{1}{e}\left[\frac{{\hat r}}{r^2} \, g\({\hat r}\) + \frac{1}{r^3} \( {\vec L}\({\hat r}\) + {\hat r}\; h\({\hat r}\)\) + \ldots \right]
\qquad \qquad \qquad r\rightarrow \infty
\lab{magneticfield}
\ee
where
\be
g\({\hat r}\)=-2 \,{\hat r}\cdot {\vec K}-\frac{i}{2}\varepsilon_{ijk}\,{\hat r}_i\sbr{K_j}{K_k}+ r\, {\vec \nabla}\cdot {\vec K}
\lab{gdef}
\ee
and
\be
h\({\hat r}\)=-3 \,{\hat r}\cdot {\vec L}-i\,\varepsilon_{ijk}\,{\hat r}_i\sbr{K_j}{L_k}+ r\, {\vec \nabla}\cdot {\vec L}
\lab{idef}
\ee
We now use the fact that the space derivatives of any function  of the radial direction only (and not of the radial distance) are of the order of $1/r$. In addition, the gradients of such functions do not have radial component. Since we are working in the gauge \rf{radialgauge} it follows that the covariant derivates of functions of ${\hat r}$ only do not have radial components either. Using such facts we can expand the covariant derivative of the Higgs field, given in  \rf{higgsatinfinity}, in powers of $1/r$ as
\be
D_i\phi= \frac{1}{r}\, \left[ r\, D_i^{(1)} \phi_0\right] + \frac{1}{r^2}\, \left[
- {\hat r}_i\, \phi_1+ r\, D^{(1)}_i \phi_1  
-i\, \varepsilon_{ijk}\, {\hat r}_j\,\sbr{L_k\({\hat r}\)}{\phi_0}\right]
+ O\(\frac{1}{r^3}\)
\lab{covderhiggsexp}
\ee
where we have denoted (see \rf{asymptoticgauge})
\be
D^{(1)}_i *\equiv \partial_i *+i\,e\, \sbr{A^{(1)}_i}{*} \qquad\qquad  {\rm with} \qquad \qquad
A^{(1)}_i=-\frac{1}{e}\, \varepsilon_{ijk}\, {\hat r}_j\,\frac{K_k\({\hat r}\)}{r}
\ee
Note that the magnetic field given in \rf{magneticfield} does not have terms of order $1/r$. Therefore, when we impose the first BPS equation in \rf{bpseq}, one gets by comparing \rf{magneticfield} and \rf{covderhiggsexp} that
\be
D_i^{(1)} \phi_0=0
\lab{covderphi01overr}
\ee
Therefore, the first non-vanishing term of the covariant derivative of $\phi_0$ is of order $1/r^2$, and so it follows the result \rf{covderphi0result}. The term of order $1/r^2$ in \rf{magneticfield} has radial components only. Consequently, by comparing \rf{magneticfield} and \rf{covderhiggsexp} one gets that the first BPS equation in \rf{bpseq} implies the relation \rf{gphi1rel} together with 
\be
D^{(1)}_i \phi_1  = i\, \varepsilon_{ijk}\, \frac{{\hat r}_j}{r}\,\sbr{L_k\({\hat r}\)}{\phi_0}
\lab{nice3}
\ee
Using the first BPS equation in \rf{bpseq} one has that
\br
\ve_{ijk}D_j B_k = \cos\gamma\, \ve_{ijk}D_j D_k \phi=\frac{i\,e}{2}\,\cos\gamma\, \;\ve_{ijk}\sbr{F_{jk}}{\phi}
=-\, i\, e\,\cos\gamma\, \sbr{B_i}{\phi}
\lab{ymeq}
\er
According to \rf{magneticfield} and \rf{asymptoticgauge}, the leading term of $\sbr{B_i}{\phi}$ is $\frac{1}{e}\frac{{\hat r}_i}{r^2}\sbr{g}{\phi_0}$, and so it is of order $1/r^2$. The leading term of $\ve_{ijk}D_j B_k$ however, is only of order $1/r^3$. Therefore, \rf{ymeq} implies that 
\be
\sbr{g}{\phi_0}=0
\lab{nice4}
\ee
and so together with \rf{gphi1rel} it implies the relation \rf{niceconsequence}. Now from \rf{covderphi01overr}, \rf{niceconsequence} and the Jacobi identity we have
\be
0=\sbr{\phi_1}{D^{(1)}_i \phi_0}= D^{(1)}_i\sbr{\phi_1}{\phi_0} - \sbr{D^{(1)}_i\phi_1}{\phi_0}=
- \sbr{D^{(1)}_i\phi_1}{\phi_0}
\ee
But together with \rf{nice3} that implies that
\be
\sbr{\sbr{{\hat r}\wedge {\vec L}}{\phi_0}}{\phi_0}=0
\lab{nice5}
\ee
Assuming that $\phi_0$ is a semisimple element of ${\cal G}$, it follows from \rf{phi0semisimple2} and \rf{nice5} that $\sbr{{\hat r}\wedge {\vec L}}{\phi_0}$ belongs to ${\rm Ker}_{{\rm adj}\phi_0}$. But again from \rf{phi0semisimple2} one sees that by definition  $\sbr{{\hat r}\wedge {\vec L}}{\phi_0}$ is an element of ${\rm Im}_{{\rm adj}\phi_0}$. Since kernel and image do not have any common element it follows that
\be
\sbr{{\hat r}\wedge {\vec L}}{\phi_0}=0
\ee
and so from \rf{nice3} one gets that
\be
D^{(1)}_i \phi_1  =0
\ee
Consequently the first non-vanishing term in the covariant derivative of $\phi_1$ has to be of order $1/r^2$ and so it follows the relation \rf{phi1covconst}. Using all such results one gets that \rf{covderhiggsexp} reduces to the form given in \rf{covderhiggsexpfinal}.

In order to evaluate the charge operator $Q$ given in \rf{charge} we have to consider, on the sphere $S^2_{\infty}$ at spatial infinity, the quantity  
\be
W^{-1}\, (\alpha F_{ij}+\beta {\widetilde F}_{ij}) \,W\,\frac{dx^{i}}{d\sigma}\frac{dx^{j}}{d\tau}= \(\alpha\, \cos \gamma+\beta \, \sin\gamma\)\, W^{-1}\,\phi_1\, W\, \Sigma + O\(\frac{1}{r}\)
\lab{chargeintegrand}
\ee
where we have used \rf{covderhiggsexpfinal}, the BPS equations \rf{bpseq}, and have denoted $\Sigma=\frac{1}{r^2}\,\ve_{kij} {\hat r}_k\, \frac{dx^{i}}{d\sigma}\frac{dx^{j}}{d\tau}$, such that $\Sigma \,d\sigma\,d\tau$ is the area element on the sphere $S^2_{\infty}$. Using the definition of the Wilson line \rf{eqforw} and \rf{phi1covconst} one gets that
\be
\frac{d\;}{d\,\sigma}\(W^{-1}\,\phi_1\, W\)= W^{-1}\,D_i\phi_1\, W\, \frac{dx^{i}}{d\sigma} \sim O\(\frac{1}{r}\)
\ee
Therefore, the quantity $W^{-1}\,\phi_1\, W$ is constant along the loops scanning the sphere $S^2_{\infty}$ at spatial infinity, and so it must equal the value of $\phi_1$ at the reference point $x_R$ where $W=\one$, i.e. $W^{-1}\,\phi_1\({\hat r}\)\, W=\phi_1\({\hat r}_R\)$, with ${\hat r}_R$ being the unit vector perpendicular to $S^2_{\infty}$ at the reference point $x_R$. So, the quantity \rf{chargeintegrand} is constant on $S^2_{\infty}$, and the relations \rf{qexponential}, \rf{finalmagneticcharge} and \rf{finalelectriccharge} follow. 

If we change the choice of the reference point to $x_R^{\prime}$, then $W_{x_R\rightarrow x_R^{\prime}}\,\phi_1\({\hat r}_R\)\, W^{-1}_{x_R\rightarrow x_R^{\prime}}=\phi_1\({\hat r}_R^{\prime}\)$, where $W_{x_R\rightarrow x_R^{\prime}}$ is the Wilson line evaluated on the curve from $x_R$ to $x_R^{\prime}$. Therefore, the eigenvalues of $\phi_1\({\hat r}_R^{\prime}\)$ and $\phi_1\({\hat r}_R\)$ are the same, and so the charges given by the eigenvalues of the operators \rf{finalmagneticcharge} and \rf{finalelectriccharge} are independent of the choice of the reference point. 

In addition,  under a general gauge transformation $A_{\mu}\rightarrow g\, A_{\mu}\,g^{-1}+\frac{i}{e}\, \partial_{\mu}g\,g^{-1}$, we have that $W^{-1}\, F_{\mu\nu}\, W\rightarrow g\(x_R\)\, W^{-1}\, F_{\mu\nu}\, W\,g^{-1}\(x_R\)$, and so $Q_{B/E}\rightarrow g\(x_R\)\,Q_{B/E} \,g^{-1}\(x_R\)$. Therefore the charges, which are the eigenvalues of $Q_B$ and $Q_E$, are invariant under any gauge transformation. 
Note however, that we have worked in the radial gauge \rf{radialgauge}, and so in the definition of $\phi_0$ and $\phi_1$ in \rf{higgsatinfinity}, we have assumed that gauge. Consequently the last equalities in \rf{finalmagneticcharge} and \rf{finalelectriccharge} are valid if one allows gauge transformations that do not depend upon the radial distance $r$. Since  the Higgs field transforms under the adjoint representation it follows that under that class of gauge transformations we have that $\phi_1\({\hat r}_R\) \rightarrow  g\(x_R\)\,\phi_1\({\hat r}_R\)\,g^{-1}\(x_R\)$, and so compatible with the transformation of $Q_B$ and $Q_E$.

\section{A couple of examples}
\label{sec:examples}
\setcounter{equation}{0}

In order to clarify even further the nature of the dynamically conserved charges \rf{charge} we consider two examples corresponding to the cases where the gauge group is  $SU(2)$ and $SU(3)$. Note that the charges \rf{charge} are conserved for any solution of the Yang-Mills equations that satisfies  the boundary conditions \rf{boundcond}. Therefore, in the $SU(2)$ case we discuss not only the BPS monopole and dyon solutions, but also the Wu-Yang and 'tHooft-Polyakov monopoles and dyons. 

\subsection{$G=SU(2)$}

The gauge field and field tensor for the Wu-Yang,  the 'tHooft-Polyakov,  and the BPS  monopoles, as well as for the corresponding dyons, all associated to the gauge group $G=SU(2)$, have the same behavior at infinity. Indeed, the gauge field and field tensor at infinity, for the dyon solutions,  are given by (see \cite{ym1} for other details of this example) 
\br
 A_i &=&   -\frac{1}{e}\,\varepsilon_{ijk}\, \frac{{\hat r}_j}{r}\, T_k
 \; ;  \qquad\qquad \qquad\qquad \qquad \qquad F_{ij}=\frac{1}{e}\,\varepsilon_{ijk}\, \frac{{\hat r}_k}{r^2}\; {\hat r}\cdot T
\lab{monopole}\\
A_0&=&\frac{M}{e}\, {\hat r}\cdot T+ \frac{\kappa}{e}\,\frac{{\hat r}\cdot T}{r}+ O(\frac{1}{r^2})
\; ;  \qquad\qquad \qquad  F_{0i} =\frac{\kappa}{e}\,\frac{{\hat r}_i}{r^2}\,{\hat r}\cdot T + O(\frac{1}{r^3})
\nonumber
\er
where  $T_i$ are the generators of the $SU(2)$ Lie algebra satisfying
\be
\sbr{T_i}{T_j}=i\,\varepsilon_{ijk}\,T_k \qquad\qquad \qquad\qquad i,j,k=1,2,3
\lab{su2}
\ee
In addition, $M$ and $\kappa$ are parameters of the solution. The next to leading terms of $A_i $ and $F_{ij}$ for the 'tHooft-Polyakov,  and the BPS dyons  are exponentially decaying. In the case of the Wu-Yang dyon, i.e. when there is no Higgs field and no symmetry breaking, the formulas \rf{monopole} are true everywhere and not only at spatial infinity. In other words, there are no terms of order $r^{-2}$ and $r^{-3}$ in $A_0$ and $F_{0i}$ respectively. Therefore, comparing with the ansatz  \rf{asymptoticgauge}, we conclude that the operator $L_k\({\hat r}\)$ vanishes in the $SU(2)$ case for single dyon and monopole solutions. By comparing \rf{monopole} with \rf{geninvsqlaw} we get that in this case
\be
g\({\hat r}\)= - {\hat r}\cdot T
\ee
Using \rf{monopole} one can explicitly check that 
\be
D_i \,{\hat r}\cdot T= \partial_i\,{\hat r}\cdot T+i\,\,e\, \sbr{A_i}{{\hat r}\cdot T}=0
\lab{rtcov}
\ee
and so $g\({\hat r}\)$ does satisfies the so-called {\em generalized inverse square law} given in \rf{geninvsqlaw}. Therefore, using \rf{eqforw} one gets that
\be
\frac{d\;}{d\sigma} \(W^{-1}\,{\hat r}\cdot T\,W\)=0\qquad\qquad \rightarrow \qquad \qquad
W^{-1}\,{\hat r}\cdot T\,W= {\hat r}_R\cdot T
\lab{rtconstant}
\ee
 where ${\hat r}_R$ is the reference point on the two sphere at infinity $S^2_{\infty}$ where the loops, scanning  it, start and end. Consequently, we observe from \rf{chargeexpand} and \rf{sstildedef} that the higher charges in that expansion are powers of the first one, not only for the BPS dyon as shown above, but also for the Wu-Yang and 'tHooft-Polyakov monopoles and dyons. Therefore, using \rf{monopole} and \rf{rtconstant} one gets that the magnetic and electric charges become (notice that the results in \rf{finalmagneticcharge} and  \rf{finalelectriccharge} are valid for BPS solutions only) 
 \br
Q_B&=& \frac{1}{4\,\pi}\,\int_{S^2_{\infty}}\, d\Sigma_i \, W^{-1}\, B_i\,W
 = -\frac{1}{e}\,{\hat r}_R\cdot T
\lab{su2magneticcharge} \\
Q_E&=& \frac{1}{4\,\pi}\,\int_{S^2_{\infty}}\, d\Sigma_i \, W^{-1}\, E_i\,W 
=\frac{\kappa}{e}\,{\hat r}_R\cdot T
\lab{su2electriccharge} 
\er
Choosing the reference point to be on the negative $z$-axis, i.e. the south pole of $S^2_{\infty}$ , we get ${\hat r}_R\cdot T=-T_3$, and so
\be
q_B=\mbox{\rm eigenvalues of $Q_B$}= \frac{1}{e}\, \(j\, ,\, j-1\, ,\, j-2\, ,\, \ldots\, ,\, -j+1\, ,\, -j\)
\ee
and similarly for the eigenvalues of $Q_E$, with $j$ being the spin (integer or half-integer) of the representation of $SU(2)$ where the eigenvalues are evaluated. Notice that such charges do satisfy the quantization condition \rf{magneticquantum} in any representation. So, the Wu-Yang,  the 'tHooft-Polyakov,  and the BPS  monopoles, as well as  the corresponding dyons, all have the same dynamically conserved charges, even though they are quite different from the topological point of view, i.e. there is no definition of  topological charge for the Wu-Yang monopole.

\subsection{$G=SU(3)$}

We now consider $SU(3)$ BPS monopole solutions, and so all the formulas discussed in sections \ref{sec:intro} and \ref{sec:calc} apply to this case. We first discuss the maximal breaking where the v.e.v. of the Higgs field is not orthogonal to any of the roots. We can take for instance ${\vec v}$ as the sum of the two simple roots, i.e. 
\be
\phi_0^{{\rm max.}} = {\vec v}\cdot {\vec h}= v\, \({\vec \alpha}_1+{\vec \alpha}_2\)\cdot {\vec h}= 
v\, {\rm diag}\(1\, ,\, 0\, ,\, -1\)
\ee
where in the last equality we have chosen the triplet representation for the Cartan sub-algebra generators $h_a$, and have normalized the roots as ${\vec \alpha}_a^2=2$, and so ${\vec \alpha}_1\cdot{\vec\alpha}_2=-1$. The gauge group $SU(3)$ is spontaneously broken to $H=U(1)\otimes U(1)$ (generated by $h_1$ and $h_2$). The topological charge of the solutions is given by the second homotopy group of the (connected) Higgs  vacua
\be
\pi_2\(SU(3)/U(1)\otimes U(1)\)= \IZ\otimes \IZ
\ee
 From \rf{topcahragecartan} we then have that the topological charge in this case is
\be
Q_B^{{\rm Top.(max.)}}= v\, \frac{\pi}{e}\, \(n_1+n_2\)
\ee
Let us now consider the case of minimum symmetry breaking where the v.e.v. of the Higgs field is orthogonal to one of the simple roots, let us say ${\vec \alpha}_1$. Then 
\be
\phi_0^{{\rm min.}} = {\vec v}\cdot {\vec h}= v\, {\vec \lambda}_2\cdot {\vec h}= 
\frac{v}{3}\, {\rm diag}\(1\, ,\, 1\, ,\, -2\)
\ee
where ${\vec \lambda}_2=\(2\,{\vec \alpha}_2+{\vec \alpha}_1\)/3$ is the second fundamental weight of $SU(3)$, and where we have used the triplet representation for $h_a$. The gauge group $SU(3)$ in this case is broken to $U(2)$, and the topological charge is determined by the homotopy group
\be
\pi_2\(SU(3)/U(2)\)= \IZ
\ee
Again from \rf{topcahragecartan} we then have that the topological charge in this case is
\be
Q_B^{{\rm Top.(min.)}}= v\, \frac{\pi}{e}\, n_2
\ee
Now, from \rf{gcartan} and \rf{omegaquant}  we  have
\be
g\({\hat r}\)= \frac{1}{2}\,\( n_1\,{\vec \alpha}_1+n_2\,{\vec \alpha}_2\)\cdot {\vec h}=
\frac{1}{2}\,{\rm diag} \(n_1\, , \, n_2-n_1\,,\,-n_2\)
\ee
where again, in the last equality, we have used the triplet representation. Therefore, from \rf{finalmagneticcharge} and \rf{dinamicalchargesexplicit}, the dynamically conserved charges are given by 
\be
q_B=\mbox{\rm eigenvalues of $Q_B$}= \frac{1}{2\,e}\,\sum_{a=1}^{2} n_a\,
\left( m_a^{(1)}\, ,\, m_a^{(2)}\, ,\,\ldots \, ,\,m_a^{(d)}\right)
\lab{dinamicalchargesexplicitsu3}
\ee
where $m_a^{(s)}$ are the eigenvalues of $2{\vec \alpha}_a\cdot {\vec h}/{\vec \alpha}_a^2$, in the representation of dimension $d$. If we choose the triplet representation, we get 
\be
q_B^{{\rm triplet}}= \frac{1}{2\,e}\,\(n_1\,,\,n_2-n_1\,,\,-n_2\)
\ee
If we had chosen the adjoint representation instead, we would get
\be
q_B^{{\rm adjoint}}= \frac{1}{2\,e}\,\(n_1+n_2\,,\,2\,n_1-n_2\,,\, -n_1+2\,n_2\,,\, 0\,,\,0\,,\, -2\,n_1+n_2\,,\, n_1-2\,n_2\,,\, -n_1-n_2\)
\ee
So, if the magnetic field has the same magnetic weights $n_a$, in the maximal and minimal symmetry breaking cases, the dynamically conserved charges are equal in both cases. So, it is insensitive to the pattern of symmetry break.  Even though the vectors $q_B$ change from one representation to another, we observe that the dynamically conserved charges correspond in fact to the magnetic weights $n_a$. Even though such weights are widely used in the literature, they were never shown to be conserved either dynamically or topologically.

\vspace{4cm}

\noindent {\bf Acknowledgements} The authors are very grateful to many helpful discussions with Peter Forgacs, Peter Goddard, Derek Harland, Betti Hartmann,  Marco Kneipp, Jutta Kunz, Nick Manton, Daniel Nogradi,  Fidel  Schaposnik, Paul Sutcliffe,  Richard Ward, Erick Weinberg and Wojtek Zakrzewski. LAF is partially supported by CNPq-Brazil. GL acknowledge the support by the International Cooperation Program CAPES-ICRANet financed by CAPES - Brazilian Federal Agency for Support and Evaluation of Graduate Education within the Ministry of Education of Brazil.


\newpage

\begin{thebibliography}{99}

\bibitem{ym1} 
  L.~A.~Ferreira and G.~Luchini, 
  ``Integral form of Yang-Mills equations and its gauge invariant conserved charges,'' 
  Phys.\ Rev.\ D {\bf 86}, 085039 (2012); 
  [arXiv:1205.2088 [hep-th]].
  %%CITATION = ARXIV:1205.2088;%%
  
\bibitem{ym2} 
  L.~A.~Ferreira and G.~Luchini, 
  ``Gauge and Integrable Theories in Loop Spaces,''
  Nucl.\ Phys.\ B {\bf 858}, 336 (2012); 
  [arXiv:1109.2606 [hep-th]].
  %%CITATION = ARXIV:1109.2606;%%
  
\bibitem{bogo} 
  E.~B.~Bogomolny, 
  ``Stability of Classical Solutions,''
  Sov.\ J.\ Nucl.\ Phys.\  {\bf 24}, 449-454 (1976)
  [Yad.\ Fiz.\  {\bf 24}, 861 (1976)].
  %%CITATION = SJNCA,24,449;%%
  
\bibitem{prasad} 
  M.~K.~Prasad and C.~M.~Sommerfield, 
   ``An Exact Classical Solution for the 't Hooft Monopole and the Julia-Zee Dyon,''
  Phys.\ Rev.\ Lett.\  {\bf 35}, 760-762 (1975).
  %%CITATION = PRLTA,35,760;%%  
  
  \bibitem{goddardolivereview} 
  P.~Goddard and D.~I.~Olive, 
  ``New Developments in the Theory of Magnetic Monopoles,''
  Rept.\ Prog.\ Phys.\  {\bf 41}, 1357 (1978).
  %%CITATION = RPPHA,41,1357;%%
  
\bibitem{mantonsut} 
  N.~S.~Manton and P.~Sutcliffe, 
  ``Topological solitons,''
  Cambridge, UK: Univ. Pr. (2004)  
  
  \bibitem{weinbergyi} 
  E.~J.~Weinberg and P.~Yi, 
  ``Magnetic Monopole Dynamics, Supersymmetry, and Duality,''
  Phys.\ Rept.\  {\bf 438}, 65-236 (2007); 
  [hep-th/0609055].
  %%CITATION = HEP-TH/0609055;%%
  
  \bibitem{weinbergbook} 
  E.~J.~Weinberg, 
  ``Classical solutions in quantum field theory : Solitons and Instantons in High Energy Physics,'' Cambridge University Press (2012).
  %%CITATION = INSPIRE-1210249;%%
  
  
  \bibitem{sutcliffereview} 
  P.~M.~Sutcliffe, 
  ``BPS monopoles,''
  Int.\ J.\ Mod.\ Phys.\ A {\bf 12}, 4663 (1997); 
  [hep-th/9707009].
  %%CITATION = HEP-TH/9707009;%%
  
  \bibitem{shnir} 
  Y.~M.~Shnir, 
  ``Magnetic Monopoles,''  Springer-Verlag (2005)
  %%CITATION = INSPIRE-1263141;%%
  
   \bibitem{nahm1} 
 W.~Nahm, ``The construction of all self-dual multimonpoles by the ADHM method'', in 
  Proceedings of {\em Monopoles In Quantum Field Theory},  eds. N.~S.~Craigie, P.~Goddard and W.~Nahm,  pags. 87-94, Trieste, Italy, December 11-15, 1981, Singapore: World Scientific (1982) 440p 

  
  \bibitem{nahm2} 
  W.~Nahm, 
  ``A Simple Formalism for the BPS Monopole,'' 
  Phys.\ Lett.\ B {\bf 90}, 413 (1980).
  %%CITATION = PHLTA,B90,413;%%
  
   
  \bibitem{nahm3} 
  W.~Nahm, 
  ``On Abelian Selfdual Multi - Monopoles,'' 
  Phys.\ Lett.\ B {\bf 93}, 42 (1980).
  %%CITATION = PHLTA,B93,42;%%
  
  \bibitem{nahm4} 
  W.~Nahm, 
  ``Selfdual Monopoles And Calorons,''
  Proceedings of the {\em  XIIth International Colloquium Held at the International Centre for Theoretical Physics} , Trieste, Italy, September 5Ð11, 1983,  pag. 189-200; preprint 
  BONN-HE-83-16, C83-09-05.1.
  %%CITATION = BONN-HE-83-16, C83-09-05.1;%%

  
  \bibitem{ward1} 
  R.~S.~Ward, 
  ``A Yang-Mills Higgs Monopole of Charge 2,''
  Commun.\ Math.\ Phys.\  {\bf 79}, 317 (1981).
  %%CITATION = CMPHA,79,317;%%
  
  \bibitem{ward2} 
  R.~S.~Ward, 
  ``Two Yang-Mills Higgs Monopoles Close Together,''
  Phys.\ Lett.\ B {\bf 102}, 136 (1981).
  %%CITATION = PHLTA,B102,136;%%
  
  \bibitem{corrigangoddard} 
  E.~Corrigan and P.~Goddard, 
  ``An $n$ Monopole Solution With 4n-1 Degrees of Freedom,''
  Commun.\ Math.\ Phys.\  {\bf 80}, 575-587 (1981).
  %%CITATION = CMPHA,80,575;%%
  
  \bibitem{prasadsinha} 
  M.~K.~Prasad, A.~Sinha and L.~L.~C.~Wang, 
  ``A Systematic Framework for Generating Multi - Monopole Solutions,''
  Phys.\ Rev.\ D {\bf 23}, 2321 (1981).
  %%CITATION = PHRVA,D23,2321;%%
  
  \bibitem{prasadrossi} 
  M.~K.~Prasad and P.~Rossi, 
  ``Construction of Exact Multimonopole Solutions,''
  Phys.\ Rev.\ D {\bf 24}, 2182-2199 (1981).
  %%CITATION = PHRVA,D24,2182;%%
  
  \bibitem{forgacs1} 
  P.~Forgacs, Z.~Horvath and L.~Palla,
  ``Exact Multi - Monopole Solutions in the Bogomolny-Prasad-Sommerfield Limit,''
  Phys.\ Lett.\ B {\bf 99}, 232 (1981)
  [errata Phys.\ Lett.\ B {\bf 101}, 457 (1981)].
  %%CITATION = PHLTA,B99,232;%%
  
  \bibitem{forgacs2} 
  P.~Forgacs, Z.~Horvath and L.~Palla, 
  ``Generating Monopoles of Arbitrary Charge by Backlund Transformations,''
  Phys.\ Lett.\ B {\bf 102}, 131 (1981).
  %%CITATION = PHLTA,B102,131;%%
  
  \bibitem{forgacs3} 
  P.~Forgacs, Z.~Horvath and L.~Palla, 
  ``Soliton Theoretic Framework for Generating Multi - Monopoles,''
  Annals Phys.\  {\bf 136}, 371 (1981).
  %%CITATION = APNYA,136,371;%%
  
  \bibitem{hitchin1} 
  N.~J.~Hitchin, 
  ``Monopoles and Geodesics,''
  Commun.\ Math.\ Phys.\  {\bf 83}, 579 (1982).
  %%CITATION = CMPHA,83,579;%%
  
  \bibitem{donaldson} 
  S.~K.~Donaldson, 
   ``Nahm's Equations and The Classification Of Monopoles,''
  Commun.\ Math.\ Phys.\  {\bf 96}, 387 (1984).
  %%CITATION = CMPHA,96,387;%%
  
  \bibitem{hurtubise1} 
  J.~Hurtubise, 
  ``Monopoles and Rational Maps: A Note on a Theorem of Donaldson,''
  Commun.\ Math.\ Phys.\  {\bf 100}, 191 (1985).
  %%CITATION = CMPHA,100,191;%%
  
   \bibitem{jutta} 
  B.~Kleihaus and J.~Kunz, 
  ``A Monopole - anti-monopole solution of the SU(2) Yang-Mills-Higgs model,''
  Phys.\ Rev.\ D {\bf 61}, 025003 (2000); 
  [hep-th/9909037].
  %%CITATION = HEP-TH/9909037;%%
  
   \bibitem{betti} 
  B.~Hartmann, B.~Kleihaus and J.~Kunz, 
  ``Dyons with axial symmetry,''
  Mod.\ Phys.\ Lett.\ A {\bf 15}, 1003 (2000); 
  [hep-th/0004108].
  %%CITATION = HEP-TH/0004108;%%
     
  \bibitem{jaffe} 
  A.~M.~Jaffe and C.~H.~Taubes, 
  ``Vortices and Monopoles: Structure of Static Gauge Theories,'' 
  Birkh\"auser, Boston, USA (1980)  (Progress in Physics, 2)
  
  \bibitem{hitchin2} 
  N.~J.~Hitchin, 
  ``On the Construction of Monopoles,''
  Commun.\ Math.\ Phys.\  {\bf 89}, 145 (1983).
  %%CITATION = CMPHA,89,145;%%
  
  \bibitem{hurtubise2} 
  J.~Hurtubise, 
   ``The Asymptotic Higgs Field of a Monopole,''
  Commun.\ Math.\ Phys.\  {\bf 97}, 381 (1985).
  %%CITATION = CMPHA,97,381;%%
  
  \bibitem{hurtubisemurray} 
  J.~Hurtubise and M.~K.~Murray, 
  ``On the Construction of Monopoles for the Classical Groups,''
  Commun.\ Math.\ Phys.\  {\bf 122}, 35 (1989).
  %%CITATION = CMPHA,122,35;%%
  
 \bibitem{nogradi} 
  D.~Nogradi, 
  ``Multi-calorons and their moduli,'' (PhD Thesis); 
  hep-th/0511125.
  %%CITATION = HEP-TH/0511125;%%
  
  \bibitem{dereknogradi} 
  D.~Harland and D.~Nogradi, to appear 
  
  \bibitem{goddardnuytsolive}  
  P.~Goddard, J.~Nuyts and D.~I.~Olive, 
  ``Gauge Theories and Magnetic Charge,''
  Nucl.\ Phys.\ B {\bf 125}, 1 (1977).
  %%CITATION = NUPHA,B125,1;%%
  
  \bibitem{englert} 
  F.~Englert and P.~Windey, 
  ``Quantization Condition for 't Hooft Monopoles in Compact Simple Lie Groups,''
  Phys.\ Rev.\ D {\bf 14}, 2728 (1976).
  %%CITATION = PHRVA,D14,2728;%%
  
  \bibitem{corriganolive} 
  E.~Corrigan and D.~I.~Olive, 
  ``Color and Magnetic Monopoles,''
  Nucl.\ Phys.\ B {\bf 110}, 237 (1976).
  %%CITATION = NUPHA,B110,237;%%
  
  \bibitem{weinberg1} 
  E.~J.~Weinberg and P.~Yi, 
  ``Explicit multimonopole solutions in SU(N) gauge theory,''
  Phys.\ Rev.\ D {\bf 58}, 046001 (1998); 
  [hep-th/9803164].
  %%CITATION = HEP-TH/9803164;%%
  
  \bibitem{weinberg2} 
  K.~M.~Lee, E.~J.~Weinberg and P.~Yi, 
  ``Massive and massless monopoles with non Abelian magnetic charges,''
  Phys.\ Rev.\ D {\bf 54}, 6351 (1996); 
  [hep-th/9605229].
  %%CITATION = HEP-TH/9605229;%%
  
  \bibitem{houghton} 
  C.~J.~Houghton, 
   ``New hyperKahler manifolds by fixing monopoles,''
  Phys.\ Rev.\ D {\bf 56}, 1220 (1997)
  [hep-th/9702161].
  %%CITATION = HEP-TH/9702161;%%
  
 \bibitem{ymoriginal} 
  C.~N.~Yang and R.~L.~Mills, 
  ``Conservation of Isotopic Spin and Isotopic Gauge Invariance,''
  Phys.\ Rev.\  {\bf 96}, 191 (1954).
  %%CITATION = PHRVA,96,191;%% 
  
   \bibitem{sweinbergbook} 
  S.~Weinberg, 
  ``The quantum theory of fields. Vol. 2: Modern applications,''
  Cambridge, UK: Univ. Pr. (1996) 
  

 \bibitem{wuyang}
 T.~T.~Wu and C.~-N.~Yang,
  ``Some solutions of the classical isotopic gauge field equations,''
  In *Yang, C.N.: Selected Papers 1945-1980*, 400-405 also in *H. Mark and S. Fernbach, Properties Of Matter Under Unusual Conditions*, New York 1969, 349-345.  
 
 \bibitem{new}
 C. P. Constantinidis, L. A. Ferreira and G. Luchini, to appear 
 
  
\end{thebibliography}
\end{document}